\documentclass[letterpaper,twocolumn,prl,aps,superscriptaddress,amsmath,amssymb,floatfix]{revtex4-1}
\usepackage{mathptmx}
\usepackage[latin9]{inputenc}
\usepackage{color}
\usepackage{amsmath}
\usepackage{amssymb}
\usepackage{graphicx}
\usepackage{esint}
\usepackage[unicode=true,
 bookmarks=true,bookmarksnumbered=false,bookmarksopen=false,
 breaklinks=false,pdfborder={0 0 1},backref=false,colorlinks=true]
 {hyperref}
\hypersetup{
 linkcolor=magenta,urlcolor=blue,citecolor=blue,pdfstartview={FitH},hyperfootnotes=false}

\makeatletter

\usepackage{textcomp}
\usepackage{epstopdf}

\usepackage{amsfonts}

\pdfpageheight\paperheight
\pdfpagewidth\paperwidth

\@ifundefined{textcolor}{}{%
 \definecolor{BLACK}{gray}{0}
 \definecolor{WHITE}{gray}{1}
 \definecolor{RED}{rgb}{1,0,0}
 \definecolor{GREEN}{rgb}{0,1,0}
 \definecolor{BLUE}{rgb}{0,0,1}
 \definecolor{CYAN}{cmyk}{1,0,0,0}
 \definecolor{MAGENTA}{cmyk}{0,1,0,0}
 \definecolor{YELLOW}{cmyk}{0,0,1,0}
}

\usepackage{xcolor}\usepackage{soul}
\definecolor{blue}{rgb}{0,0,1}
\definecolor{red}{rgb}{1,0,0}
\definecolor{green}{rgb}{0,1,0}

\makeatother

\begin{document}
\title{Noiseless photonic non-reciprocity via optically-induced magnetization}
\affiliation{CAS Key Laboratory of Quantum Information, University of Science and
Technology of China, Hefei 230026, P. R. China.}
\affiliation{State Key Laboratory of Quantum Optics and Quantum Optics Devices,
and Institute of Opto-Electronics, Shanxi University, Taiyuan 030006,
China}
\author{Xin-Xin Hu}
\thanks{These two authors contributed equally to this work.}
\affiliation{CAS Key Laboratory of Quantum Information, University of Science and
Technology of China, Hefei 230026, P. R. China.}
\affiliation{CAS Center For Excellence in Quantum Information and Quantum Physics,
University of Science and Technology of China, Hefei, Anhui 230026,
P. R. China.}
\author{Zhu-Bo Wang}
\thanks{These two authors contributed equally to this work.}
\affiliation{CAS Key Laboratory of Quantum Information, University of Science and
Technology of China, Hefei 230026, P. R. China.}
\affiliation{CAS Center For Excellence in Quantum Information and Quantum Physics,
University of Science and Technology of China, Hefei, Anhui 230026,
P. R. China.}
\author{Pengfei Zhang}
\email{zhangpengfei@sxu.edu.cn}

\affiliation{State Key Laboratory of Quantum Optics and Quantum Optics Devices,
and Institute of Opto-Electronics, Shanxi University, Taiyuan 030006,
China}
\affiliation{Collaborative Innovation Center of Extreme Optics, Shanxi University,
Taiyuan 030006, China}
\author{Guang-Jie Chen}
\affiliation{CAS Key Laboratory of Quantum Information, University of Science and
Technology of China, Hefei 230026, P. R. China.}
\affiliation{CAS Center For Excellence in Quantum Information and Quantum Physics,
University of Science and Technology of China, Hefei, Anhui 230026,
P. R. China.}
\author{Yan-Lei Zhang}
\affiliation{CAS Key Laboratory of Quantum Information, University of Science and
Technology of China, Hefei 230026, P. R. China.}
\affiliation{CAS Center For Excellence in Quantum Information and Quantum Physics,
University of Science and Technology of China, Hefei, Anhui 230026,
P. R. China.}
\author{Gang Li}
\affiliation{State Key Laboratory of Quantum Optics and Quantum Optics Devices,
and Institute of Opto-Electronics, Shanxi University, Taiyuan 030006,
China}
\affiliation{Collaborative Innovation Center of Extreme Optics, Shanxi University,
Taiyuan 030006, China}
\author{Xu-Bo Zou}
\affiliation{CAS Key Laboratory of Quantum Information, University of Science and
Technology of China, Hefei 230026, P. R. China.}
\affiliation{CAS Center For Excellence in Quantum Information and Quantum Physics,
University of Science and Technology of China, Hefei, Anhui 230026,
P. R. China.}
\author{Tiancai Zhang}
\affiliation{State Key Laboratory of Quantum Optics and Quantum Optics Devices,
and Institute of Opto-Electronics, Shanxi University, Taiyuan 030006,
China}
\affiliation{Collaborative Innovation Center of Extreme Optics, Shanxi University,
Taiyuan 030006, China}
\author{Hong X. Tang}
\affiliation{Deparment of Electric Engineering, Yale University, New Haven, CT
06511, USA}
\author{Chun-Hua Dong}
\email{chunhua@ustc.edu.cn}

\affiliation{CAS Key Laboratory of Quantum Information, University of Science and
Technology of China, Hefei 230026, P. R. China.}
\affiliation{CAS Center For Excellence in Quantum Information and Quantum Physics,
University of Science and Technology of China, Hefei, Anhui 230026,
P. R. China.}
\author{Guang-Can Guo}
\affiliation{CAS Key Laboratory of Quantum Information, University of Science and
Technology of China, Hefei 230026, P. R. China.}
\affiliation{CAS Center For Excellence in Quantum Information and Quantum Physics,
University of Science and Technology of China, Hefei, Anhui 230026,
P. R. China.}
\author{Chang-Ling Zou}
\email{clzou321@ustc.edu.cn}

\affiliation{CAS Key Laboratory of Quantum Information, University of Science and
Technology of China, Hefei 230026, P. R. China.}
\affiliation{CAS Center For Excellence in Quantum Information and Quantum Physics,
University of Science and Technology of China, Hefei, Anhui 230026,
P. R. China.}
\affiliation{State Key Laboratory of Quantum Optics and Quantum Optics Devices,
and Institute of Opto-Electronics, Shanxi University, Taiyuan 030006,
China}
\date{\today}
\begin{abstract}
The realization of optical non-reciprocity is crucial for many device
applications, and also of fundamental importance for manipulating
and protecting the photons with desired time-reversal symmetry. Recently,
various new mechanisms of magnetic-free non-reciprocity have been
proposed and implemented, avoiding the limitation of the strong magnetic
field imposed by the Faraday effect. However, due to the difficulties
in suppressing the drive and its induced noises, these devices exhibit
limited isolation performances and leave the quantum noise properties
rarely studied. Here, we demonstrate a new approach of magnetic-free
non-reciprocity by optically-induced magnetization in an atom ensemble.
Excellent isolation of signal (highest isolation ratio is $51.4_{-2.5}^{+6.5}\,\mathrm{dB}$)
is observed over a power dynamic range of $7$ orders of magnitude,
with the noiseless property verified by quantum statistics measurement.
The approach is applicable to other atoms and atom-like emitters in
solids, paving the way for future studies of integrated photonic non-reciprocal
devices, unidirectional quantum storage and state transfer, as well
as topological photonics technologies. 
\end{abstract}
\maketitle

\section{Introduction}

\noindent Lorentz reciprocity and its violation, associating with
the time-reversal symmetry breaking, are of fundamental and conceptual
importance in optics~\cite{Potton2004,Teich2007,Asadchy2020}, and
have also led to controversies in the photonics research community~\cite{Jalas2013}.
In practical optical applications, non-reciprocal optical devices,
including isolator, circulator, and gyrator, are indispensable and
ubiquitous. The prominent mechanisms realizing the optical non-reciprocity
are the magnetic circular dichroism and circular birefringence (Faraday
effect) in bulky magneto-optical materials~\cite{Stephens1974,Kaminsky2000}.
A strong magnetic bias field changes the dipole momentum or transition
frequencies of dielectrics by inducing the Zeeman splitting of electron
spin states and modifying the electronic wavefunctions, thus breaks
the reciprocity of light~\cite{Asadchy2020}. However, limited by
magnetic shields, material processing, switching response and reconfigurability,
the utilization of conventional non-reciprocal devices is restricted
in many scenarios, such as photonic integrated circuits~\cite{Dtsch2005,Bi2011}
and hybrid superconducting-photonic systems~\cite{Fan2018,Elshaari2020}.

Over the past decade, great efforts have been dedicated to realizing
magnetic-free non-reciprocity of light~\cite{Asadchy2020}. Ingenious
ideas and new experimental techniques are developed to break the limit
of the conventional Faraday effect, such as optical drive induced
directional frequency conversion~\cite{Shen2016,Ruesink2016,Shen2017,Verhagen2017},
storage~\cite{Dong2015,Kim2015,Zhang2018} or amplification~\cite{Hua2016,Li2019,Xia2018,Lin2019},
the synthetic magnetic field in a loop of coupled-resonator~\cite{Fang2017,Aidelsburger2018},
and RF/acoustic drive induced spatio-temporal modulation of refraction
index~\cite{Yu2009,Fang_2012,Sounas2017,Sohn2018}. The common principle
behind these approaches is the orbital momentum conservation in the
coherent mode conversions, which induces absorption or phase shift
for photons input from a selected direction. However, such mechanism
imposes limitations for non-reciprocal devices, such as the stringent
phase-matching condition, drive-induced noises via the intermediate
excitations or nonlinear wave-mixing processes, giving rise to non-ideal
performances. For example, the quantum properties of the isolators
are not tested experimentally, and the achievable isolation ratio
is less than 20~dB~\cite{Shen2017,Shen2016,Zhang2018,Hua2016,Bi2011,Ruesink2016,Kim2015,Li2019}.
\begin{figure}
\centerline{\includegraphics[clip,width=1\columnwidth]{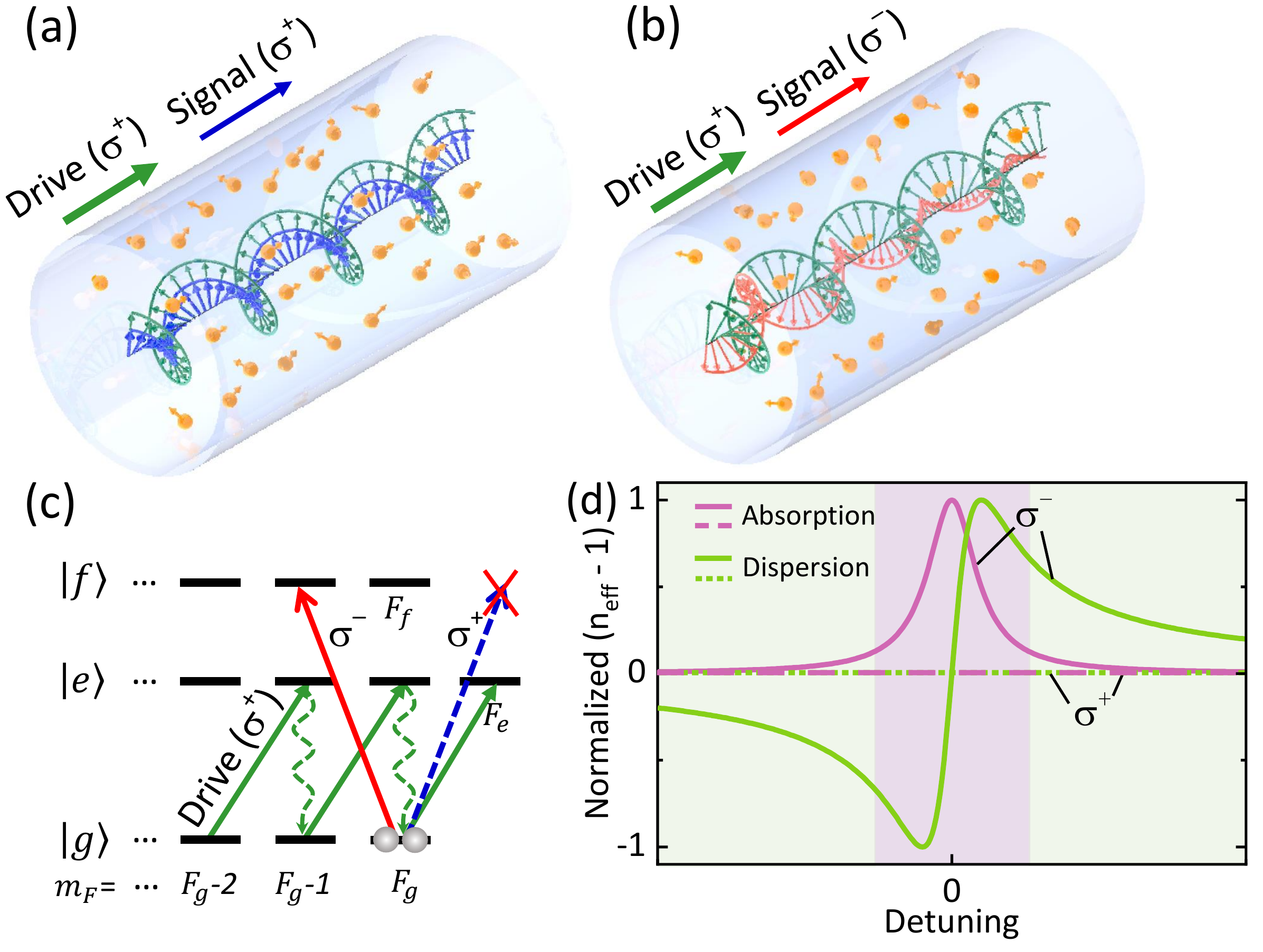}}\caption{\textbf{Schematic of circular birefringence and circular dichroism
arising from optically-induced magnetization.} (a) and (b) In an atomic
medium with non-zero magnetization induced by a $\sigma^{+}$-polarized
drive field, the $\sigma^{+}$- and $\sigma^{-}$-polarized signals
propagate at different velocities and experience different absorption
losses. (c) The energy diagram of typical atoms or atomic-like emitters,
with degenerate ground Zeeman levels ($\left|g\right\rangle $) and
the allowing optical transitions to excited energy levels ($\left|e\right\rangle $,
$\left|f\right\rangle $). The drive field ($\left|g\right\rangle \rightarrow\left|e\right\rangle $)
induces the magnetization of the atoms as the population concentrates
to one side of the Zeeman levels, which leads to the different response
to $\sigma^{+}$- and $\sigma^{-}$-polarized signals ($\left|g\right\rangle \rightarrow\left|f\right\rangle $)
due to selection rules. (d) The illustration of absorption and dispersion
in a magnetized atomic medium for $\sigma^{+}$- and $\sigma^{-}$-polarized
signals. Absorption dominates in the shadow region where the signal
is near-resonant to the atomic transitions, while dispersion dominates
for off-resonant cases.}

\label{Fig1} 
\end{figure}

Here, we propose and demonstrate a new scheme to realize the non-reciprocity
via the optically-induced magnetization (OIM) of an atomic medium.
Isolation ratios of $30.3_{-0.2}^{+0.3}\,\mathrm{dB}$ and $51.5_{-2.5}^{+6.5}\,\mathrm{dB}$
are realized with a warm atom vapor in free space and in a traveling-wave
cavity, respectively. Dispersive non-reciprocal effects analogous
to Faraday effect are also observed with a non-reciprocal optical
mode frequency splitting exceeding $100\,\mathrm{MHz}$. It is worth
noting that the non-reciprocal photon-atom interaction was investigated
previously~\cite{Sayrin2015,Scheucher2016}, however, under the requirement
of tens of Gauss bias magnetic field. The magnetic-free OIM mechanism
is distinct by treating the atom ensemble as a magnetizable dielectric
medium, where incoherent population transfer could be utilized for
achieving the non-reciprocity. Its crucial advantages include that
the optical drive frequency could be far detuned from the signal,
avoiding the difficulty associated with phase matching and drive filtration,
robustness against drive fluctuations, and elimination of drive-induced
noises at signal frequencies. We show that the equivalent intracavity
noise excitation introduced by the device is less than $10^{-2}$,
proving the inherent noiseless property of this mechanism. We believe
that our experiments could stimulate further experimental and theoretical
efforts on OIM, to demonstrate non-reciprocal devices in photonic
integrated circuits, and find applications in quantum photonic chips
and explore intriguing topological properties of light~\cite{Ozawa2019}.

\vbox{}

\section{Results}

\subsection{Principle of OIM-based non-reciprocity}

\noindent Figures~\ref{Fig1}(a) and (b) schematically illustrate
the general principle of the noiseless all-optical non-reciprocity
that is based on the OIM. In the presence of an external circularly-polarized
drive field, the optical response of an atom ensemble to a weak signal
field is modified, leading to a circular-polarization dependent propagation
velocity and absorption. As indicated in Fig.~\ref{Fig1}(c), the
atoms possess hyperfine ground spin states $\left|g,m_{F}\right\rangle $,
with $m_{F}$ denoting the quantum spin number with $\left|m_{F}\right|\leq F_{g}$.
By introducing the ancillary energy level $\left|e\right\rangle $
with $F_{e}\geq F_{g}$, the drive laser with the $\sigma^{+}$ polarization
changes the population of the atoms to $m_{F}=F_{g}$, and thus builds
up an effective magnetization of the atom spin states (green lines
in Fig.~\ref{Fig1}(c)). For an input signal near-resonant to $\left|g\right\rangle \rightarrow\left|f\right\rangle $
by $F_{f}\leq F_{g}$, the drive laser could transfer the population
to $m_{F}=F_{g}$, for which the its $\sigma^{+}$-transition is forbidden
(blue dash line in Fig.~\ref{Fig1}(c)). Therefore, the atomic medium
in Fig.~\ref{Fig1}(a) is transparent to the $\sigma^{+}$-polarized
signal. In contrast, the transition for the $\sigma^{-}$-polarized
signal is allowed (red solid line in Fig.~\ref{Fig1}(c) ), thus
the signal propagation is described in the atomic media configuration
shown in Fig.~\ref{Fig1}(b).

Figure~\ref{Fig1}(d) illustrates the OIM-induced circular birefringence
and dichroism, as the polarized medium induces velocity and absorption
change to the probe signals. When the circularly polarized probe signal
is off-resonant to the atoms (with detuning $\Delta$), the atoms
induce effective phase change $\propto1/\Delta$ while the absorption
is suppressed to $\propto1/\Delta^{2}$. Different from the conventional
Faraday effect, where the external magnetic field induces the Zeeman
energy level shifts and thereby circular birefringence, our scheme
bases on ground state magnetization of the atoms induced by circularly
polarized drive. The scheme is also distinct from previous nonlinear
optical schemes~\cite{Shen2016,Ruesink2016,Shen2017,Verhagen2017,Dong2015,Kim2015,Zhang2018,Hua2016,Li2019,Xia2018,Lin2019},
where the phase-matching condition breaks the time-reversal symmetry
of light. The drive field could be applied to any ancillary transition
that couples to the target ground levels and is not necessary to be
coherent with the signal. Therefore, the scheme allows broad bandwidth
non-reciprocity beyond the limitation of phase-matching, permits more
convenient filtration of drive laser, and the device would not induce
noises for signals. It is also noted that the population condition
illustrated in Fig.~\ref{Fig1}(c) is not strictly required for realizing
non-reciprocity, because the time-reversal symmetry could be achieved
as long as the uniform population distribution over all $m_{F}$ states
is broken to produce nonzero net spin polarization. 
\begin{figure*}
\centerline{\includegraphics[clip,width=0.8\textwidth]{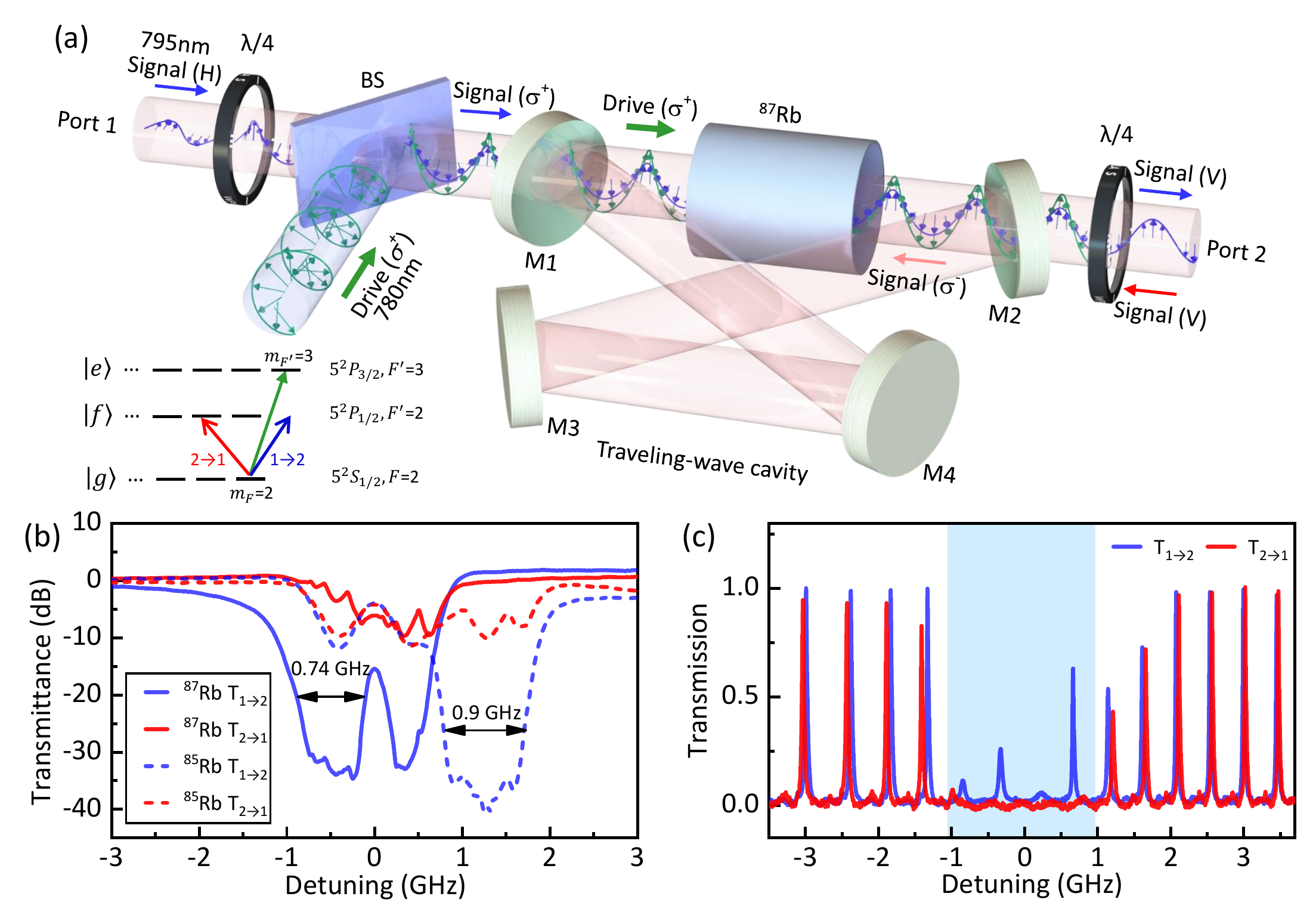}}\caption{\textbf{Experimental observation of the magnetic-free non-reciprocity
in an atom ensemble.} (a) Experimental setup with a traveling-wave
cavity, comprising of four mirrors with a Rb vapor cell inside. The
linearly polarized input signal from port 1(2) is converted to $\sigma^{+}$($\sigma^{-}$)
polarization by quarter waveplates (QWP). A drive laser in $\sigma^{+}$
polarization is coupled into the cavity through a beam splitter of
port 1. Therefore, the circular birefringence or dichroism leads to
non-reciprocal transmission of the whole device system. The inset
diagram shows the energy levels of Rb atom, with the arrows representing
the drive laser (green), the forward signal laser from port 1 (blue)
and the backward signal laser from port 2 (red). (b) The broadband
isolation by hot atom ensemble due to optically-induced magnetization,
based on the setup in (a) without the cavity. A maximum isolation
ratio of $30.3_{-0.2}^{+0.3}\,\mathrm{dB}$ and $29.6_{-0.6}^{+0.6}\,\mathrm{dB}$
is observed for $^{87}\mathrm{Rb}$ and $^{85}\mathrm{Rb}$, respectively,
and the corresponding $20\,\mathrm{dB}$-isolation bandwidth are $740\,\mathrm{MHz}$
and $900\,\mathrm{MHz}$. 
Here, the drive power ($P_{\mathrm{d}}$) is about $130\,\mathrm{mW}$
while the signal power ($P_{\mathrm{s}}$) is about $1\,\mathrm{\mu W}$.
(c) Typical transmission spectra for the cavity setup, with $P_{\mathrm{d}}=45\,\mathrm{mW}$
and $P_{\mathrm{s}}=1\,\mathrm{\mu W}$. The blue shadow indicates
the absorptive region that exhibits circular dichroism, and the region
outside the shadow shows mode splitting due to circular birefringence.}

\label{Fig2} 
\end{figure*}

We experimentally implement the proposed scheme based on $^{87}\mathrm{Rb}$
warm atom vapor, as shown in Fig.~\ref{Fig2}(a). The energy diagram
of the atom is shown in the inset, with $\left|g\right\rangle =\left|5^{2}S_{1/2},F=2\right\rangle $
and $\left|f\right\rangle =\left|5^{2}P_{1/2},F'=1,2\right\rangle $
for the $D_{1}$ transitions at $\sim795\,\mathrm{nm}$, and we choose
the ancillary energy levels $\left|e\right\rangle =\left|5^{2}P_{3/2},F'=1,2,3\right\rangle $
corresponding to $D_{2}$ transitions at $\sim780\,\mathrm{nm}$.
First of all, the non-reciprocity via OIM is verified by a circularly
polarized drive field ($\sim780\,\mathrm{nm}$) on the atoms, with
$\sigma^{+}$-polarized forward (port $1\rightarrow2$) or $\sigma^{-}$-polarized
backward (port $2\rightarrow1$) signal probing the vapor cell at
$\sim795\,\mathrm{nm}$. Here, the predicted circular dichroism of
atomic medium is converted to non-reciprocal transmittance ($T_{1\rightarrow2}\neq T_{2\rightarrow1}$)
of linear polarization introduced by way of a pair of quarter waveplates
(QWP)
 (Fig.~\ref{Fig2}(a)). As plotted in Fig.~\ref{Fig2}(b), we observe
a $20\,\mathrm{dB}$ isolation with bandwidth of about $740\,\mathrm{MHz}$
due to the Doppler broadening from the atom transition linewidth of
$\sim6\,\mathrm{MHz}$. The highest isolation ratio of $30.3_{-0.2}^{+0.3}\,\mathrm{dB}$
is achieved without a cavity in a single pass configuration (details
in Supplementary Information). To valid the mechanism for other atomic
media, we also demonstrate the cavity-less isolation by the isotope
$^{85}\mathrm{Rb}$ in a vapor cell of natural abundance Rubidium,
demonstrating a $900\,\mathrm{MHz}$ bandwidth and a highest isolation
ratio of $29.6_{-0.6}^{+0.6}\,\mathrm{dB}$. Since the magnetization
of the medium could be perturbed by the stray magnetic field perpendicular
to the direction of light propagation ($B_{\perp}$), the OIM-induced
non-reciprocity might be sensitive to magnetic field. However, in
our experiments, we found that the isolation is very robust and could
be maintained even with $B_{\perp}=10\,$Gauss. Therefore, the OIM
provides a very robust photonic non-reciprocity against experimental
imperfections, and also relaxes the requirements of drive laser and
operating environment for practical applications.

\vbox{}

\subsection{Absorptive and dispersive non-reciprocity.}

\noindent Since the OIM-induced non-reciprocity strongly depends on
the optical depth of the atom medium, which might be a limiting factor
for cold atoms or emitters in solids, we investigate the cavity-enhanced
non-reciprocity in the following experiments (Fig.~\ref{Fig2}(a)).
Typical spectra of the cavity-enhanced non-reciprocity are shown in
Fig.~\ref{Fig2}(c), where the shadow section corresponds to the
absorptive region in Fig.~\ref{Fig1}(d). Here, the transmission
of $T_{2\rightarrow1}$ is greatly suppressed, while peaks with regular
free-spectral range are observed in $T_{1\rightarrow2}$ transmission.
Since the atoms in the vapor cell fly between the reservoir and cavity
mode fields, which induces the relaxation of the OIM, the peaks in
the absorption region show reduced transmittance due to the residual
populations on the ground states $m_{F}<2$ ($F_{g}=2$). In contrast,
in the detuned frequency regions, the system shows resonance shift
between forward and backward spectra. These observations confirm the
absorptive and dispersive non-reciprocity realized by the OIM in the
atomic medium (Fig.~\ref{Fig1}(d)). 
\begin{figure}[tp]
\centerline{\includegraphics[clip,width=1\columnwidth]{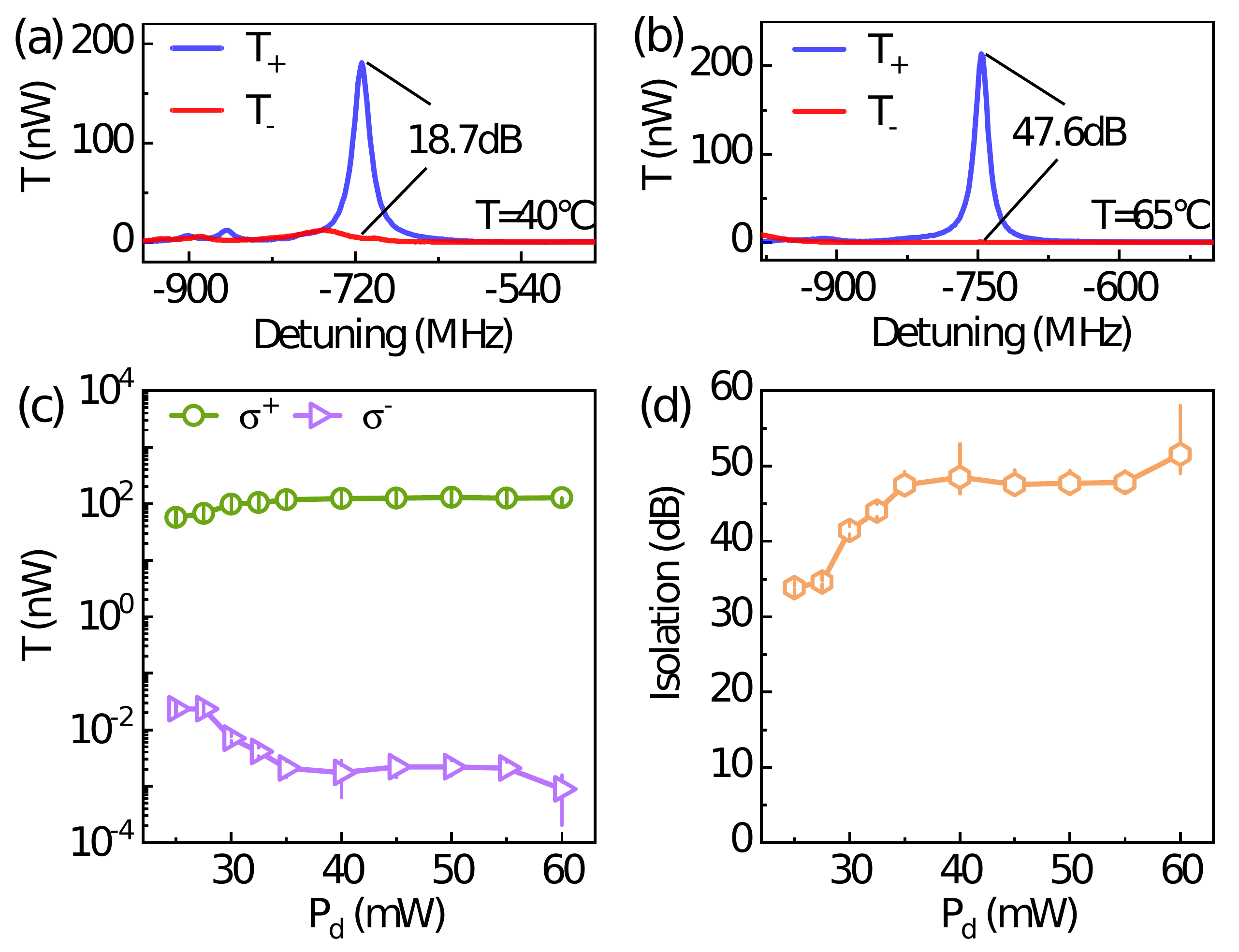}}\caption{\textbf{Optical isolation via absorptive non-reciprocity.} (a,b) Typical
transmission spectra for $\sigma^{+}$ and $\sigma^{-}$-polarized
signals at different atomic densities, with the temperature of the
atomic cell respectively set at $40\,\mathrm{{^\circ}C}$ and $65\,\mathrm{{^\circ}C}$
and the corresponding isolation ratios $18.7\,\mathrm{dB}$ and $47.6\,\mathrm{dB}$,
respectively. Here, the signal and drive power is fixed at $P_{\mathrm{s}}=50\,\mathrm{\mu W}$
and $P_{\mathrm{d}}=45\,\mathrm{mW}$, respectively. (c) The on-resonant
transmitted signal power \textit{vs} the drive laser power ($P_{\mathrm{d}}$).
The green and purple spots represent the $\sigma^{+}$ and $\sigma^{-}$-polarized
signals, respectively. (d) Extracted isolation ratio at different
$P_{\mathrm{d}}$. The largest isolation ratio $51.5_{-2.5}^{+6.5}\,\mathrm{dB}$
is achieved at $P_{\mathrm{d}}=60\,\mathrm{mW}$. The results in (c)
and (d) are measured at $65\,^{\circ}\mathrm{C}$ with a fixed signal
laser power of $P_{\mathrm{s}}=50\,\mathrm{\mu W}$, and error bars
denote standard deviations.}
\label{Fig3} 
\end{figure}

In an optical traveling-wave cavity, optical modes of orthogonal polarizations
are degenerated. As demonstrated above, the OIM could induce the circular
birefringence and circular dichroism and thus lifts the degeneracy
between $\sigma^{+}$- and $\sigma^{-}$-polarized modes in our system.
Then, the non-reciprocity is realized by probing the OIM via quarter
waveplates, as the signal transmission for port $2\rightarrow1$ with
$\sigma^{-}$-polarization is rejected by the cavity. For better calibration
of the system performance, we probe the $\sigma^{-}$-polarization
from the forward direction by utilizing the symmetry of the system,
because all optical elements are the same when measuring both $\sigma^{+}$
and $\sigma^{-}$ polarization forwardly ($T_{+}$ and $T_{-}$) except
that the angle of waveplate is different. 
Detailed performances of the optical isolation are summarized in Fig.~\ref{Fig3}.
By increasing the temperature of the cell (Fig.~\ref{Fig3}(a) and
(b)), the isolation ratio increases due to the increase of atom density
under a fixed drive power $P_{\mathrm{d}}$ of 45mW. The optical performance
also depends on the drive power. As shown in Fig.~\ref{Fig3}(c),
the $T_{+}$ is suppressed at low $P_{\mathrm{d}}$, because the OIM
is weak and the atom ensemble is absorptive for both polarizations.
By increasing $P_{\mathrm{d}}$, the atoms are effectively magnetized
and become transparent to the $\sigma^{+}$-polarized signal. Figure~\ref{Fig3}(d)
shows the extracted isolation ratio, which increases with $P_{\mathrm{d}}$
and reaches a maximum of \textcolor{black}{$51.5_{-2.5}^{+6.5}\,\mathrm{dB}$}
when $P_{\mathrm{d}}=60\,\mathrm{mW}$. The linewidth here at $65\,^{\circ}\mathrm{C}$
is around $26.5_{-0.2}^{+0.2}\,\mathrm{MHz}$, primarily determined
by the glass-cell-induced loss. 
\begin{figure}[tp]
\centerline{\includegraphics[clip,width=1\columnwidth]{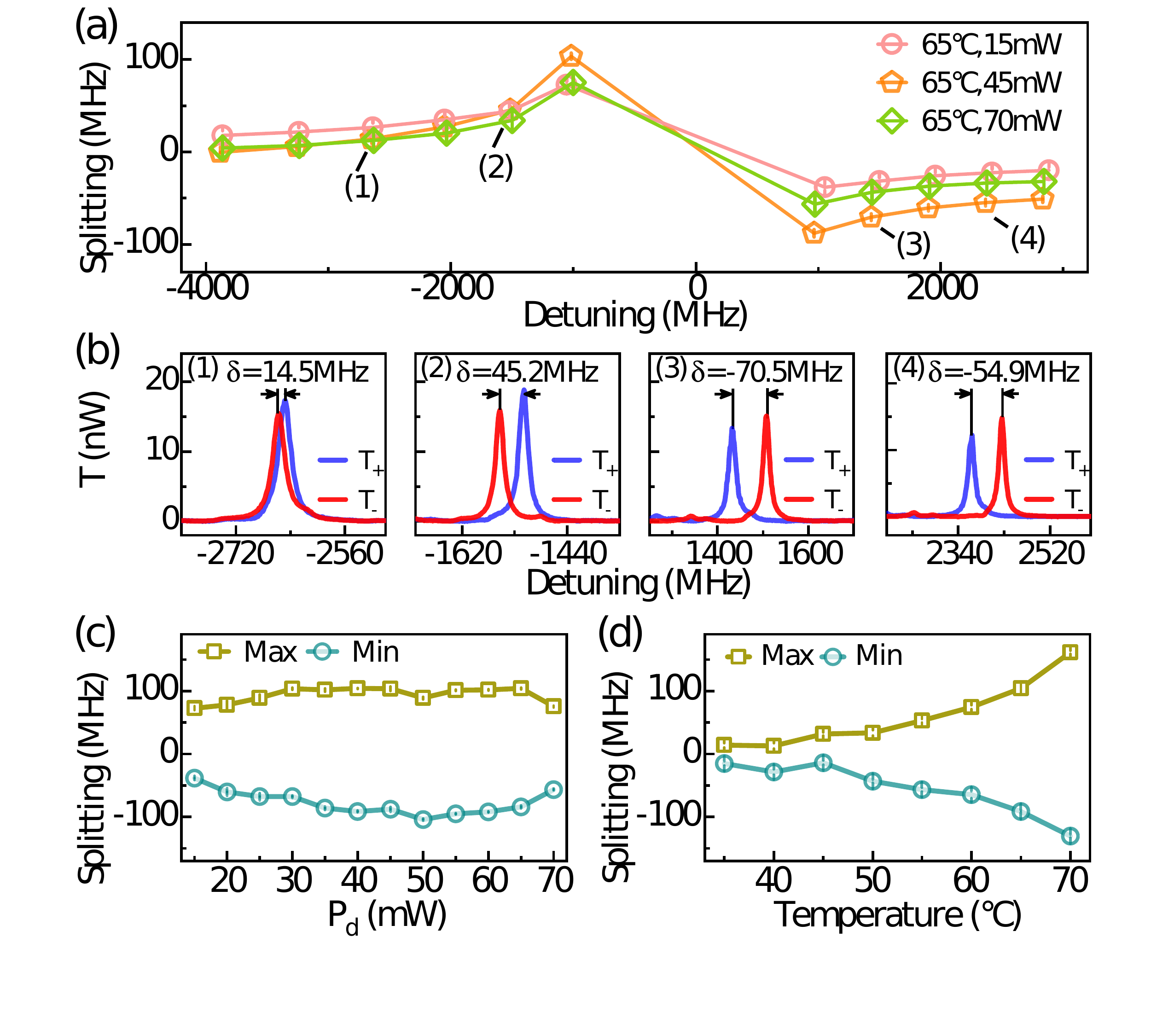}}\caption{\textbf{Mode frequency shift due to dispersive non-reciprocity.} 
(a) The resonant frequency shift of two circularly polarized modes
at different drive laser power levels ($P_{\mathrm{d}}$) and fixed
cell temperature of $65\,\mathrm{^{\circ}C}$. When the frequency
of signal is near-resonant to the atoms (blue shadow region in Fig.
2c), the absorptive effect dominates and the resonances for $T_{-}$
are suppressed. The dispersive shift can not be distinguished in this
regime. (b) The relative mode spectral shifts at four different cavity
detunings labeled in (a). (c) and (d) The extracted maximum and minimum
mode frequency shifts against $P_{\mathrm{d}}$ and different cell
temperatures. In all studies, the signal laser power is fixed at $P_{\mathrm{s}}=20\,\mathrm{\mu W}$,
and error bars denote standard deviations.}

\label{Fig4} 
\end{figure}

Figure~\ref{Fig4} presents the results of the dispersive non-reciprocal
effects. The resonance shift $\delta=f_{\mathrm{+}}-f_{\mathrm{-}}$
of the $T_{+}$ and $T_{-}$ spectra at different temperature are
shown in Fig.~\ref{Fig4}(a), where $f_{+(-)}$ is the resonant frequency
of $T_{+}$ ($T_{-}$) spectrum. Figure~\ref{Fig4}(b) further details
the transmission spectra corresponding to four detuning frequencies
indicated in Fig.~\ref{Fig4}(a). At $65\,^{\circ}\mathrm{C}$ and
$P_{\mathrm{d}}=45\,\mathrm{mW}$, we find that the ratio between
mode splitting and linewidth $\delta/\gamma$ reaches\textcolor{black}{{}
$4.55$}, where $\gamma$ denotes the linewidth of the cavity at $f_{\mathrm{+}}$.
The $\delta/\gamma$ is a figure-of-merit characterizing OIM for isolator
and gyrator to account for non-reciprocal phase shift induced by OIM.
Furthermore, we find that the relative frequency shift monotonously
increases with the vapor temperature as shown in Fig.~\ref{Fig4}(d)
due to the increase of atom density at elevated temperatures. For
the power dependence shown in Fig.~\ref{Fig4}(c), the frequency
shift initially increases and then drops when $P_{\mathrm{d}}$ approaches
$70\,\mathrm{mW}$ due to population transfer from atomic ground states
to the ancillary energy levels ($\left|f,m_{F^{'}}=3\right\rangle $)
at high drive intensities, since the drive is coupling with a cyclic
transition ($\left|g,m_{F}=2\right\rangle \rightarrow\left|f,m_{F^{'}}=3\right\rangle $).
\begin{figure}[tp]
\centerline{\includegraphics[clip,width=1\columnwidth]{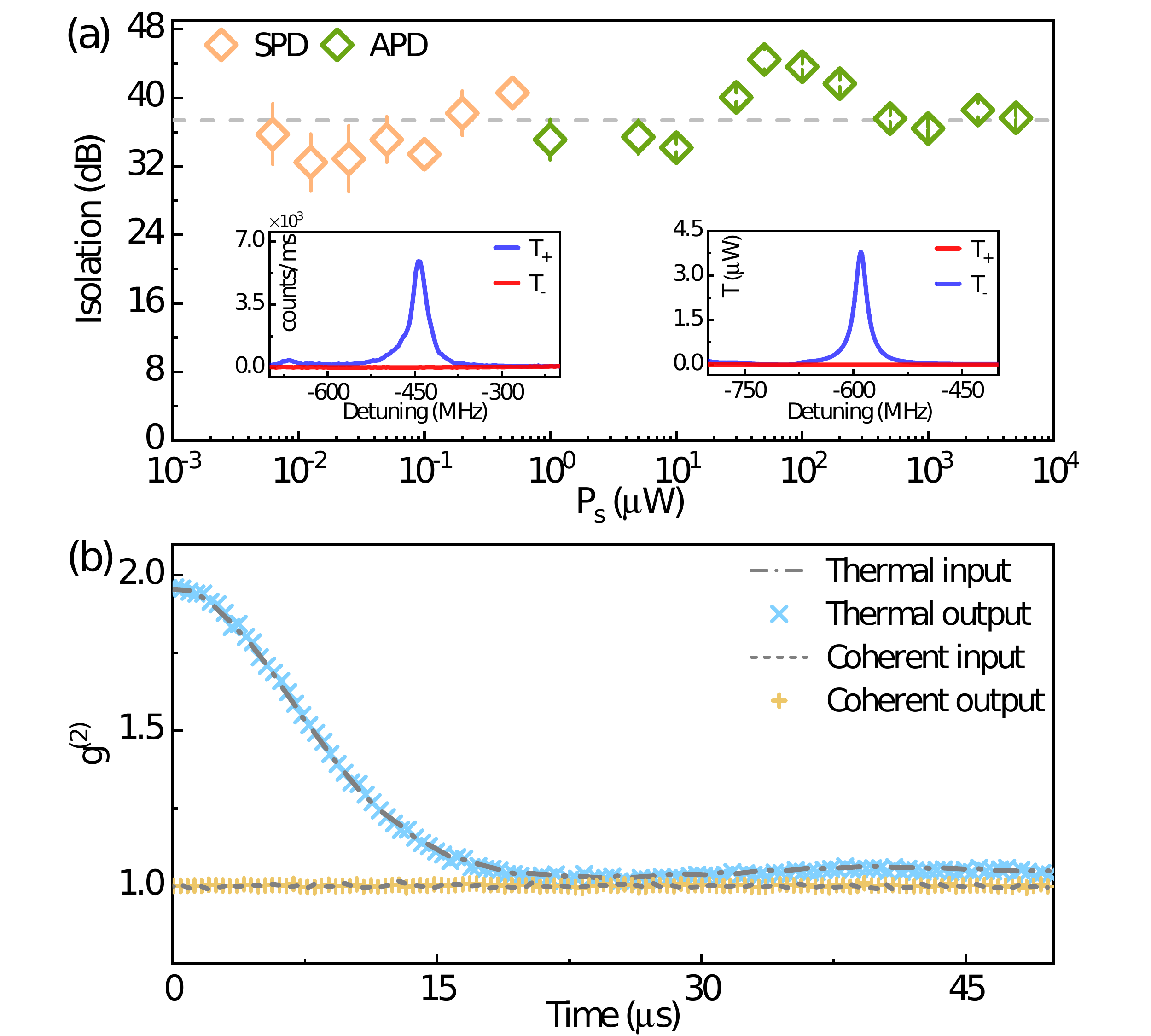}}\caption{\textbf{The dynamic range and noiseless property of the all-optical
isolation.} (a) The measured isolation ratios against input signal
power ($P_{\mathrm{s}}$). Due to the limitation of the system insertion
loss and detector's response, avalanche photodetectors (APDs) are
adapted for $P_{\mathrm{s}}\protect\geq1\,\mathrm{\mu W}$ (shown
as green spots) and a single-photon detector (SPD) is used for $P_{\mathrm{s}}<1\,\mathrm{\mu W}$
(shown as orange spots). 
The insets are two typical spectra measured for $P_{s}=6.25\,\mathrm{nW}$
(left) and $1.2\,\mathrm{mW}$ (right), respectively. (b) Measured
second-order correlation function ($\mathrm{g^{(2)}}$) for signal
input to and output from the device. Blue crosses and dashed line
are the results for pseudo-thermal source, while yellow crosses and
dotted line are the results for coherent source. In all studies, the
drive power is set at $50\,\mathrm{mW}$, vapor temperature is $65\,\mathrm{{^\circ}C}$,
and error bars denote standard deviations.}
\label{Fig5} 
\end{figure}

\vbox{}

\subsection{Noiseless isolation}

\noindent We lastly demonstrate the inherent noiseless property of
our all-optical non-reciprocity via OIM. In the previous demonstration
of all-optical non-reciprocity, the strong drive would induce considerable
noise by four-wave mixing amplification~\cite{Hua2016,Lin2019,Li2019}
or the thermal noise occupation of low-frequency excitations~\cite{Shen2016,Fang2017}.
In contrast, there are no such noise processes in our approach that
could excite the $D_{1}$ transitions by the drive at room temperature,
thus there should be no noise photon generated at signal wavelengths.
We verify this characteristic over a wide signal power range varying
from nano- to milli-Watt, corresponding to an intracavity photon number
in the range of $0.84-10^{6}$. As shown in Fig.~\ref{Fig5}(a),
the isolator shows an average of about $37.4\,\mathrm{dB}$ isolation
with a signal power dynamic range of about $70\,\mathrm{dB}$. Fluctuations
of the performance are attributed to the varied response of the detectors
at different input power, such as the saturated gain of avalanche
photodetector (APD) and counts saturation of single-photon detector
(SPD). To quantify the quantum noise properties of OIM, we measure
the second-order correlation function ($g^{(2)}\left(t\right)$ with
a delay time $t$) of transmitted signals, since a noiseless photonic
device should conserve the quantum statistics of the input signal.
As presented in Fig.~\ref{Fig5}(b), the $g^{(2)}\left(t\right)$
is measured for both coherent laser and pseudo-thermal light source
(see Method and Supplementary Information for details). By comparing
the $g^{(2)}\left(t\right)$ curves for input and output signal, we
find the isolator conserves the quantum statistics of these signals.
Assuming possible stray thermal and coherent photons detected by SPD,
we estimate the equivalent added intracavity noise photon number is
$n_{\mathrm{add}}\sim0.0084$ (See Supplementary Materials for details),
which confirms the noiseless non-reciprocity mechanism proposed in
this work.

\vbox{}

\section{Discussion}

\noindent Comparing to previously demonstrated all-optical non-reciprocity
originated from coherent nonlinear processes, the optically-induced
atomic magnetization is an incoherent process, thus is very robust
against experimental imperfections, such as the drive frequency and
amplitude fluctuations, inhomogeneous transition frequency broadening,
imperfect circular polarization and beam alignment, as well as stray
magnetic fields. The demonstrated noiseless non-reciprocity only requires
an atomic ensemble with degenerate Zeeman energy levels, thus the
mechanism is extendable to many atoms or atom-like emitters, including
hot and cold atoms, molecules, as well as emitters in solids (such
as NV centers and rare-earth atoms). Thus, the high-performance isolation
is achievable even in UV and mid-IR wavelengths, where commercial
products have very limited performance. On a photonic chip, the optically-induced
magnetization could be conveniently implemented by harnessing the
spin-orbit coupling of photon~\cite{Sayrin2015,Sollner2015,Lodahl2017}.
Furthermore, our approach could be generalized to other particles
or quasi-particles other than photons, such as phonons by utilizing
their interaction with electron spins in solids~\cite{Golter2016}.
Aside from the potential applications, our work also merits new physics
by taking the degenerate ground Zeeman energy levels into cavity quantum
electrodynamics (QED)~\cite{Yang2019} and waveguide QED, where interesting
non-reciprocal phenomena arise, such as non-reciprocal multi-stability,
quantum frequency conversion, photon storage and lasing.

\bibliographystyle{Zou}
\bibliography{AtomIsolator}

\vbox{}

\noindent \textbf{Acknowledgments}\\
 We would like to thank Shu-Hao Wu and Ji-Zhe Zhang for the assistance
in collecting and processing data, Ming Li and Xin-Biao Xu for helpful
discussions. The work was supported by the National Key R\&D Program
of China (Grant No.~2016YFA0301303), the National Natural Science
Foundation of China (Grant No. 11922411, 11874342, 11704370, 11874342,
and 91536219), and Anhui Initiative in Quantum Information Technologies
(Grant No.~AHY130200). P. Z, G. L. and T. Z. were supported by National
Key R\&D Program of China (Grant No. 2017YFA0304502), the National
Natural Science Foundation of China (Grants No. 11974225, No. 11574187,
No. 11674203, No. 11974223, and No. 11634008), and the Fund for Shanxi
``1331 Project'' Key Subjects Construction. C.-L.Z. was also supported
by the Program of State Key Laboratory of Quantum Optics and Quantum
Optics Devices (No.~KF201809).

\vbox{}

\noindent \textbf{Author contributions}\\
 C.-L.Z., X.-X.H., G.L. and H.T. conceived the experiments. X.-X.H.,
Z.-B.W. and P.Z. built the experimental setup and carried out the
measurements with the assistance by G.-J.C., G.L. and T.Z. X.-X.H.,
Z.-B.W. and C.-L.Z. analyzed the data, Y.-L.Z. and X.-B.Z. provided
theoretical supports. C.-L.Z., X.-X.H. and P.Z. wrote the manuscript
with input from all co-authors. C.-H.D., G.-C.G. and C.-L.Z. supervised
the project. All authors contributed extensively to the work presented
in this paper.
\end{document}